\documentclass[amsmath,amssymb,aps,10pt,prd,letterpaper,nofootinbib,balancelastpage,notitlepage,superscriptaddress,twocolumn,floatfix]{revtex4-2}
\usepackage{graphicx}	
\usepackage{epstopdf}	
\usepackage{dcolumn}	
\usepackage{ragged2e}	
\usepackage{bm}		    
\usepackage{enumitem}   
\usepackage{rotating}
\usepackage[sort&compress]{natbib}	 	
	\setcitestyle{square,numbers,comma}	
\usepackage[colorlinks=true,urlcolor=blue,linkcolor=black,citecolor=red]{hyperref}	

\newcommand{\orcid}[1]{\href{https://orcid.org/#1}{\,\includegraphics[width=8px]{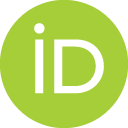}}}

\newcommand{\tabref}[2][]{Tab{#1}.~\ref{#2}}		
\newcommand{\sectref}[2][]{Sec{#1}.~\ref{#2}}		
\newcommand{\appref}[2][x]{Appendi{#1}~\ref{#2}}	
\renewcommand{\eqref}[2][]{Eq{#1}.~(\ref{#2})}		
\newcommand{\citeR}[2][]{Ref{#1}.~\cite{#2}}			

\renewcommand{\Re}{\ensuremath\,\mathrm{Re}}

\newcommand{\lb}{\ensuremath{\left}}					
\newcommand{\rb}{\ensuremath{\right}}					
\newcommand{\nl}{\nonumber \\ & \quad }					

\newcommand{\order}[1]{\ensuremath{\mathcal{O}(#1)}}    
\newcommand{\Hz}[1]{\ensuremath{\,\mathrm{{#1}Hz}}}

\newcommand{\gw}{\textsc{gw}}
\newcommand{\WD}{\textsc{wd}}

\begin{document}

\title{Astrometric Gravitational-Wave Detection via Stellar Interferometry}
\date{\today}

\author{Michael A.~Fedderke\orcid{0000-0002-1319-1622}}
\email{mfedderke@jhu.edu}
\affiliation{The William H.~Miller III Department of Physics and Astronomy, The Johns Hopkins University, Baltimore, MD  21218, USA}
\author{Peter W.~Graham\orcid{0000-0002-1600-1601}}
\email{pwgraham@stanford.edu}
\affiliation{Stanford Institute for Theoretical Physics, Department of Physics, Stanford University, Stanford, CA 94305, USA}
\affiliation{Kavli Institute for Particle Astrophysics \& Cosmology, Department of Physics, Stanford University, Stanford, CA 94305, USA}
\author{Bruce Macintosh\orcid{0000-0003-1212-7538}}
\email{bmacint@stanford.edu}
\affiliation{Kavli Institute for Particle Astrophysics \& Cosmology, Department of Physics, Stanford University, Stanford, CA 94305, USA}
\author{Surjeet Rajendran\orcid{0000-0001-9915-3573}\,}
\email{srajend4@jhu.edu}
\affiliation{The William H.~Miller III Department of Physics and Astronomy, The Johns Hopkins University, Baltimore, MD  21218, USA}

\begin{abstract}
We evaluate the potential for gravitational-wave (GW) detection in the frequency band from 10\,nHz to 1\,$\mu$Hz using extremely high-precision astrometry of a small number of stars.  
In particular, we argue that non-magnetic, photometrically stable, hot white dwarfs (WD) located at $\sim$ kpc distances may be optimal targets for this approach.
Previous studies of astrometric GW detection have focused on the potential for less precise surveys of large numbers of stars; our work provides an alternative optimization approach to this problem.
Interesting GW sources in this band are expected at characteristic strains around $h_c \sim 10^{-17} \times \left(\mu \text{Hz} / f_{\gw}\right)$.
The astrometric angular precision required to see these sources is $\Delta \theta \sim h_c$ after integrating for a time $T \sim 1/f_{\gw}$.
We show that jitter in the photometric center of WD of this type due to starspots is bounded to be small enough to permit this high-precision, small-$N$ approach.  
We discuss possible noise arising from stellar reflex motion induced by orbiting objects and show how it can be mitigated.
The only plausible technology able to achieve the requisite astrometric precision is a space-based stellar interferometer.
Such a future mission with few-meter-scale collecting dishes and baselines of $\mathcal{O}(100\,\text{km})$ is sufficient to achieve the target precision.
This collector size is broadly in line with the collectors proposed for some formation-flown, space-based astrometer or optical synthetic-aperture imaging-array concepts proposed for other science reasons.
The proposed baseline is however somewhat larger than the km-scale baselines discussed for those concepts, but we see no fundamental technical obstacle to utilizing such baselines.
A mission of this type thus also holds the promise of being one of the few ways to access interesting GW sources in this band.
\end{abstract}

\maketitle

\tableofcontents

\section{Introduction}
\label{sect:intro}
The universal nature of gravitation implies that every object in the Universe emits gravitational waves~(GW). 
Further, these waves travel unhindered through the Universe, carrying information about the physics of their production across the aeons of space and time. 
Gravitational waves thus enable unique probes of the Universe. 
They are the only known way to probe the near-horizon geometry of black holes and the physics of the early Universe prior to recombination. 
Compact astrophysical objects such as white dwarfs (WD), neutron stars, and black holes that are faint or dark in the electromagnetic spectrum are also expected to be copious producers of gravitational waves. 
The historic detection of gravitational waves by the LIGO/Virgo Collaborations~\cite{PhysRevLett.116.061102,PhysRevLett.119.161101,PhysRevX.9.031040,Abbott:2020niy,LIGOScientific:2021djp} has made it possible for us to begin to explore this rich physics, and an interesting anomaly at much lower frequencies in pulsar timing array data~\cite{Arzoumanian:2020vkk,Goncharov:2021oub,2021MNRAS.508.4970C,2022MNRAS.510.4873A} may potentially be the first hint of a new discovery just around the corner.

Terrestrial optical-interferometer detectors such as LIGO/Virgo/KAGRA are sensitive to gravitational waves above $\sim 10$\,Hz~\cite{PhysRevLett.116.061102,PhysRevLett.119.161101,PhysRevX.9.031040,Abbott:2020niy,LIGOScientific:2021djp,Akutsu:2020zlw,LIGOScientific:2022myk}. 
Since many astrophysical processes occur at frequencies lower than 10\,Hz, there is a strong science case to detect gravitational waves at lower frequencies~\cite{Sesana:2019vho,Baibhav:2019rsa,Sedda:2019uro,Baker:2019pnp}. 
There are a number of active experimental efforts to achieve this goal.
These include pulsar timing arrays (PTAs)~\cite{Kramer_2013,Babak:2015lua,Shannon1522,2016MNRAS.458.1267V,Aggarwal:2018mgp,Kerr_2020,Arzoumanian:2020vkk,Goncharov:2021oub,2021MNRAS.508.4970C,2022MNRAS.510.4873A} that operate around nHz--$\mu$Hz; the LISA constellation~\cite{Baker:2019nia,LISA_Sci_Req,LISA_L3} that is aimed at 1--10\,mHz; TianQin aimed at 0.01--1\Hz{}~\cite{Luo:2015ght,Milyukov:2020fm}; atomic-interferometry approaches such as MAGIS/MIGA/AION/AEDGE/ZAIGA~\cite{Dimopoulos:2007cj,Dimopoulos:2008sv,Hogan:2010fz,Graham:2017pmn,Coleman:2018ozp,Canuel:2018fq,Abe:2021ksx,Tino:2019tkb,Badurina:2019hst,Zhan_2019,AEDGE:2019nxb,Badurina:2021rgt} around 1\,Hz; clock-based proposals~\cite{Kolkowitz:2016wyg} between mHz and Hz; DECIGO at 0.1--10\,Hz~\cite{Kawamura:2018esd,kawamura2020current}; and Cosmic Explorer~\cite{CosmicExplorer} and the Einstein Telescope~\cite{Maggiore:2019uih} above $\sim 10$\,Hz. 
Concepts have also been developed to detect gravitational waves in the $\mu$Hz--mHz band using LISA-style constellations~\cite{Sesana:2019vho}, using asteroids as test masses in a future space-based mission~\cite{Fedderke:2021kuy}, studying orbital perturbations to various binary systems~\cite{Blas:2021mqw,Blas:2021mpc}, and looking for low-frequency modulation of higher-frequency GWs~\cite{Bustamante-Rosell:2021daj}.
Existing astrometric studies (e.g., \citeR[s]{Pyne:1995iy,schutz_2009,Book:2010pf,Klioner:2017asb,Moore:2017ity,Wang:2020pmf,Wang:2022sxn,2018FrASS...5...11V}) also access this band, as we discuss in more detail below.
Additionally, \citeR{Park:2019ies} proposed a concept to access part of the band below LISA by using the interference of starlight from a single star collected by dishes on separate formation-flown satellites, in order to monitor the GW-induced fluctuations in the proper distance between local TMs housed in the satellites.

Preliminary work has also been performed~\cite{Sesana:2019vho} to identify the robust science case for gravitational-wave detection in the frequency band $10\,\text{nHz} \lesssim f_{\gw} \lesssim \mu\text{Hz}$ (see also \citeR{Spallicci:2011nr}). 
Very roughly, \citeR{Sesana:2019vho} establishes that in this frequency band, a detector would need to be sensitive to characteristic strains as small as $h_c \sim 10^{-17} \times \left( \mu \text{Hz} / f_{\gw} \right)$ in order to successfully detect expected astrophysical signals, such as (super)massive black-hole binary mergers.
Most existing proposals in the 10\,nHz--$\mu$Hz band do not have sufficient sensitivity to access this level of strain (although see \citeR[s]{Sesana:2019vho,Fedderke:2021kuy} for mission concepts that may be able to achieve this sensitivity).

Gravitational waves cause fluctuations in the space-time between inertial test masses, and they can thus be detected by observing these fluctuations. 
Due to the smallness of the expected signals, these test masses have to be highly immune to environmental noise. 
This requirement is particularly difficult at low frequencies. 
One way to tackle this problem is to engineer environmental isolation systems for local test masses.
This is the approach that has been taken by the LISA constellation~\cite{LISA_L3}, and the LISA Pathfinder mission~\cite{PhysRevLett.120.061101} has successfully demonstrated this key technological element in the LISA band. 
However, it has recently been shown~\cite{Fedderke:2020yfy} that in the frequency band 10\,nHz--$\mu$Hz, gravity gradient noise (GGN) arising from the large population of inner Solar System asteroids acting on local test masses placed within the inner Solar System is significantly larger than the expected gravitational-wave signal.%
\footnote{\label{ftnt:bothEnds}%
    This noise estimate is applicable for approaches where \emph{both} TMs (i.e., both ends of the baseline) are located in the (inner) Solar System.} %
This asteroid GGN cannot be shielded, as it arises from a gravitational coupling to the test masses.
Moreover, it is potentially significant even in the outer edges of the Solar System at the lowest end of that band. 
There is thus a strong need for alternative detection approaches in this frequency band. 

Distant astrophysical objects may serve as natural inertial test masses if their intrinsic astrophysical properties are sufficiently stable. 
In an approximate sense, this approach has been pursued in pulsar timing arrays, which exploit the extreme rotational stability of pulsars, and where the distant nature of the pulsar helps to overcome terrestrial backgrounds (including gravity gradient noise~\cite{PhysRevD.30.732}) that inhibit the use of local test masses below $\sim1$\,Hz.
Such distant test masses, being outside the inner Solar System, are also not subject to the same asteroid GGN limitation that will plague other approaches at these frequencies.
The limitation of this approach is however the nature and abundance of such astrophysical objects:%
\footnote{\label{ftnt:nyquist}%
    Were all pulsars to be sampled with temporally evenly spaced observations, the high-frequency cutoff in PTA sensitivity would set by the corresponding Nyquist frequency, which would typically be in the $\sim \mu$Hz range for observations performed at $\sim 1$--2 week intervals. 
    However, somewhat higher frequency coverage is achieved in practice by higher-cadence sampling of a subset of pulsars (see, e.g., \citeR{Perera:2018pts}); alternatively, appropriately staggered low-cadence sampling also allows access to higher frequencies (see, e.g., \citeR{Wang:2020hfh}). 
    These subtleties notwithstanding, the important points are that PTAs are most sensitive around $f_{\gw} \sim T_{\text{obs}}^{-1} \sim 1$--$3\,\text{nHz}$, with $T_{\text{obs}} \sim (1$--$3)\times 10^1\,$yrs~\cite{Arzoumanian:2020vkk,Goncharov:2021oub,2021MNRAS.508.4970C,2022MNRAS.510.4873A} being the duration of the timing campaign, and that they lose sensitivity rising through the frequency band of interest to this work.} %
the population and characteristics of millisecond pulsars in the vicinity of the Earth limits the sensitivity of pulsar timing arrays above 10\,nHz. 
We thus need a different kind of astrophysical test mass. 

Another interesting way to detect gravitational waves in this frequency band is via stellar astrometry (i.e., measurements of the positions of stars). 
In this approach, the star itself serves directly as an inertial test mass, and the gravitational wave causes the relative angular separation of multiple stars to fluctuate. 
Using stellar astrometry in this way to detect gravitational waves is not a new idea; in fact, it has been well explored: see, e.g., \citeR[s]{Pyne:1995iy,schutz_2009,Book:2010pf,Klioner:2017asb,Moore:2017ity,Wang:2020pmf,Wang:2022sxn,2018FrASS...5...11V}. 
Previous studies have however been concerned with the usage of existing (or proposed, but not dedicated) instruments to make measurements of the angles between large numbers of typical (e.g., main sequence) stars in large-scale stellar surveys (e.g., \emph{Gaia}~\cite{GaiaOverview} or future Roman Space Telescope~\cite{RomanWebsite} surveys~\cite{Wang:2020pmf}). 
However, the astrometric precision achievable in such survey data results in GW strain sensitivities for these instruments that are typically above the levels necessary to detect astrophysically relevant gravitational-wave amplitudes in the 10\,nHz--$\mu$Hz frequency band~\cite{Wang:2020pmf}. 

In this paper, we investigate an underexplored corner of the astrometric GW detection trade space. 
We ask instead whether extremely high-precision observations of a small number of stars can overcome some of the limitations of previous lower-precision%
\footnote{\label{ftnt:notLowPrecision}%
    Note that we use the word `lower' here advisedly: existing world-leading measurements and instruments are exquisitely precise; they are not however precise enough to detect GW-induced fluctuations at expected amplitudes.} %
studies of larger numbers of stars. 
In particular, we study a key question that arises when considering a smaller number of stars: what kinds of stars are sufficiently good test masses for the purpose of small-$N$ astrometric detection of astrophysically relevant gravitational-wave strains in this frequency band?
Specifically, can backgrounds arising from the presence of starspots and planetary or asteroidal bodies be sufficiently ameliorated to allow GW detection at the relevant strains?

We demonstrate that photometrically stable, non-magnetic white dwarfs (NWMD) at~$\sim$\,kpc distances from the Earth are an excellent class of astrophysical test masses with which to pursue the high-precision, small-$N$ approach. 
We show that starspot backgrounds are sufficiently suppressed for this class of stars, and discuss the prospects for (and mitigation of) any planetary backgrounds.

Having identified such NMWD as a promising class of stellar test masses, we proceed to evaluate the properties required for an instrument to be capable of achieving the necessary precision on the measurement of the angle between two or more NMWD that are at~$\sim$\,kpc distances.
The only feasible technology capable of achieving the required precision is space-based stellar interferometry, in which light from the same star is collected by multiple collecting dishes separated by a known baseline and interfered in order to detect minute changes in angular position.
Our proposed measurement would require simultaneous monitoring of multiple stars in a similar fashion.

Numerous studies of such space-based interferometers were completed in the first decade of the 2000s, although none has been realized; see \tabref{tab:spaceInterferometers} and \citeR[s]{SPECS_Spirit,Mather:1998zp,SIMLite,SIMLiteProgRep,Goullioud:2008sq,TPF-I,Darwin1,Darwin2,StellarImagerVisionMissionStudy,StellarImagerSummary,StellarImagerWebsite}.
These varied in their purpose as astrometers (i.e., instruments that measure the locations of unresolved sources) or (conventional or nulling) synthetic-aperture imagers (i.e., instruments that can form a resolved image with a resolution equivalent to a large-aperture instrument, by sparse sampling).

\begin{table*}
\begin{ruledtabular}
\caption{\label{tab:spaceInterferometers}%
    A selection of space-based interferometric mission concepts that have been proposed (mostly in the period 2000--2010). 
    We roughly categorize these missions by their purpose as either astrometers `A', or synthetic aperture imagers `I'.
    We note approximate figures for the baseline and individual collector apertures that were discussed for each concept, the number of `science' collectors proposed (marked with a `*' if an additional `combiner' hub satellite was envisaged), and the part of the spectrum the mission was to operate in.
    We also note the technology proposed to construct and stabilize the baseline: `B' for boom, referring to a rigid structure on which multiple light-collectors were to be mounted; `T' for tethered formation, referring to separate spacecraft that are loosely tethered together and formation-flown; and `F' for free formation, referring to completely independent spacecraft flown in formation.
    Many of the properties we quote here are rough or representative of values that were considered in design studies, and we refer the reader to the noted references for further detailed discussion; none of the mission concepts noted here have to date been pursued beyond the design and development stages.
    }%
\begin{tabular}{llllllll}
Mission name    &   Purpose & Typical baseline [m] & Aperture [m] & Collectors & Spectrum & Baseline technology & Refs. \\ \hline
SPIRIT  &   I   &  30--50   &   1--3    & 2     &   far IR  &   B   & \cite{SPECS_Spirit} \\
SPECS   &   I   &  1000     &   3--10   & 2--3* &   far IR  &   T   & \cite{SPECS_Spirit,Mather:1998zp}  \\
SIMS    &   I/A &  10       &   0.3     & 7     &   optical &   B   & \cite{SIMS}  \\
SIM Lite&   A   &  6        &   0.5     & 2     &   optical &   B   & \cite{SIMLite,SIMLiteProgRep,Goullioud:2008sq} \\
TPF-I/Darwin& I &  200--500 &   2--4    & 4*    &   mid-IR  &   F   & \cite{TPF-I,Darwin1,Darwin2} \\
SI Pathfinder&I &  20--50   &   1       & 3--5  &   UV      &   B/F & \cite{StellarImagerVisionMissionStudy,StellarImagerSummary,StellarImagerWebsite} \\
Stellar Imager (SI)
            & I & 500--1000 &   1--2    & 20--30*  & UV/Optical & F & \cite{StellarImagerVisionMissionStudy,StellarImagerSummary,StellarImagerWebsite}
\end{tabular}
\end{ruledtabular}
\end{table*}

To complete our measurement at the required accuracy, we find that we would need a blend of the properties that have been proposed in various prior mission studies; our proposed measurement thus constitutes a further, GW-science motivation for further development of technology for space-based stellar interferometers of multiple types.
Since relative angular fluctuation is an astrometic measurement, we would operate in a mode similar to the SIMS~\cite{SIMS} or SIM Lite~\cite{SIMLite} mission concepts, requiring only a small number of light collectors and baselines.
However, we would need to be able to simultaneously track the relative motion of multiple widely separated stars, so the mission would require at least four `science' collectors (two independent baselines).
Moreover, to image the faint and hot WD, we would need a UV imaging system with collecting apertures in the few-meter class, which brings the optical system properties to those of the Stellar Imager (SI) proposal~\cite{StellarImagerVisionMissionStudy}.
Finally, to achieve the requisite astrometric precision, we would require precision formation-flown baselines in the range of a hundred kilometers, which exceeds the largest, km-scale baselines considered in proposals such as SPECS~\cite{SPECS_Spirit,Mather:1998zp} and SI by around two orders of magnitude.
Because the mission we would need is quite similar to SI in many crucial design aspects, but requiring few collectors and needing them to be spaced further apart, it is plausible that a mission of our proposed type could serve as one kind of pathfinder or technology demonstrator mission for a future space-based imaging array akin to SI.
Ultra-high precision astrometry of this type would also enable a host of other science goals, in addition to allowing GW detection; see; e.g., \citeR[s]{Gottesman:2011eg,Nomerotski:2020adi,Stankus:2020hbc,Chakrabarti:2022hdw}.

The rest of this paper is organized as follows. 
In \sectref{sect:signal}, we describe the expected astrometric gravitational-wave signal. 
In \sectref{sect:backgrounds}, we estimate astrophysical backgrounds for this astrometric detection approach, and show that NMWD at $\sim$\,kpc distances suppress them. 
Using these inputs, we describe the necessary instrument in \sectref{sect:instrument}. 
We discuss possible alternative measurement schemes and stellar targets in \sectref{sect:otherTargets}.
We conclude in \sectref{sect:conclusions}.
The \hyperlink{app:fullSignal}{Appendix} contains detailed expressions for GW-induced astrometric deflection.

\section{The Signal}
\label{sect:signal}
A gravitational wave (with $+$ polarization) that is moving along the $z$ direction is described in transverse-traceless gauge by the metric~\cite{Misner:1974qy}: 
\begin{align}
    ds^2 = - dt^2 + &\big[1 + h_{+}^{(0)} \sin\big(\omega_{\gw}\! \left(t - z\right)\! \big)\big] dx^2 \nonumber \\ + &\big[1 - h_{+}^{(0)} \sin\big(\omega_{\gw}\! \left(t - z\right)\! \big)\big] dy^2 + dz^2.
    \label{eq:GWmetric}
\end{align}
where $h_{+}^{(0)}$ is the amplitude of the gravitational wave and $\omega_{\gw}$ its (angular) frequency: $\omega_{\gw}\equiv2\pi f_{\gw}$.
The signal of such a gravitational wave in astrometric detectors has been computed many times (see, e.g., \citeR[s]{Pyne:1995iy,schutz_2009,Book:2010pf} and references therein), with a particularly clear and careful derivation given in \citeR{Book:2010pf}.
For the reader's convenience, we reproduce the final result of \citeR{Book:2010pf} in \appref{app:fullSignal}, and summarize here only the salient features of the result.

Suppose we have a hypothetical detector that measures the angular position of a star with respect to a perfectly known reference frame. 
A gravitational wave of strain $h$ that passes through this detector will cause the measured location of the star to deflect by an angle $\Delta \theta \sim h$, with the angular deflection modulating at the frequency $f_{\gw}$ of the gravitational wave. 
The detection of this modulated deflection constitutes the detection of a gravitational wave. 
This result is valid as long as the wavelength of the gravitational wave is smaller than the distance between the detector and the star: $\lambda_{\gw} \ll d_{\WD}$ (distant source limit). 
Given that the typical distance between a detector located in the Solar System and a distant star is at least tens of parsecs ($d_{\WD} \gtrsim 3\times 10^{17}\,\text{m}$, corresponding to a light-travel time of $\gtrsim 30\,\text{yrs}$), the practical low-frequency cutoff of an astrometric GW observatory is provided by the observation time ($\sim 10\,\text{yrs}$ for a typical space-based mission) rather than the decrease in the signal amplitude that occurs once the source distance is inside the GW wavelength.
Similarly, the practical high-frequency cutoff is set by the sampling frequency (technically, the Nyquist frequency) of the observations, which is in turn limited by the flux of photons from the star, in the sense that at least one photon must be received on average during a measurement.%
\footnote{\label{eq:tradeoffOnTime}%
    Of course, except for requiring at least one photon on average is received during a measurement and also that the measurement time is not a substantial fraction of a GW period, one can choose to trade off between higher-cadence measurements at lower precision, or lower-cadence measurements at higher precision, without changing the overall sensitivity.
    That is, suppose it takes a time $\tau$ to make a single measurement of an angle to a precision $\sigma_{\theta} = \sigma_{\theta}^{(0)}/\sqrt{N}$, where $N\geq 1$ is the number of photons received in time $\tau$, and $\sigma_{\theta}^{(0)}$ is the intrinsic single-photon angular measurement accuracy (set by, e.g., the photon wavelength and optical system parameters).
    In a time $T$, one can make $M \sim T/\tau$ such measurements, resulting in an ultimate sensitivity of $\sigma_{\theta}^{\text{final}} \sim \sigma_{\theta} / \sqrt{M} = \sigma_{\theta}^{(0)} / \sqrt{M\times N}$.
    But $M\times N$ is just the total number of photons received in time $T$.
    Assuming that both $2\pi f_\gw \tau \ll 1$ and $N \gtrsim 1$, one is free to adjust $\tau$ (and thus $N$) and $M$, keeping $N \times M$ fixed, and the overall sensitivity is unchanged.
    } %
In the interval between these two frequencies, the angular deflection (i.e., the GW signal) is $\Delta \theta \sim h_c$, flat in frequency $f_{\gw}$ (of course, for any one particular source, the signal is peaked at the frequency of the source generating the GW; we mean here that the signal amplitude in the detector does not depend on that frequency for a fixed GW strain amplitude).  
 
The quadrupolar nature of the gravitational wave gives rise to a distinct pattern in the deflections of stellar positions across the sky. 
This signal is dominated by the value of the gravitational wave at the position of the detector: the `Earth term' [see \appref{app:fullSignal} and compare \eqref[s]{eq:fullSignal} and (\ref{eq:fullSignalFarSource})].
There is, in principle, an additional contribution that arises from the value of the gravitational wave at the position of the star itself. 
But this contribution is suppressed in the distant source limit, is delayed in time (i.e., it measures the amplitude of the strain at the location of the star when the light was first emitted), and does not appear as a coherent quadrupolar signal across the sky since its value will change significantly from one stellar source location to the next. 
Because the GW signal driven by the Earth term is coherent among all stars observed, but differs in amplitude and direction for each individual star, we can search for the signal as a coherent oscillation in the \emph{relative} angular positions of stars on the sky, dispensing finally with the fiction of a perfect background reference frame against which any one star is measured.
 
To detect astrophysically relevant gravitational waves with characteristic strains $h_c \sim 10^{-17} \times \left( \mu \text{Hz} / f_{\gw} \right)$, we thus need an instrument with the ability to measure relative angular changes between widely separated stars at the level of $\Delta \theta \sim 10^{-17} \times \left( \mu \text{Hz} / f_{\gw} \right)$ over the measurement band of interest; i.e., frequencies $10\,\text{nHz}\lesssim f_{\gw} \lesssim \mu\text{Hz}$.

\section{Astrophysical Backgrounds} 
\label{sect:backgrounds}
Successful detection of gravitational waves using small numbers of stars requires the astrometrically measured relative angular position of the individual stars to fluctuate due to backgrounds by no more than an amount of order the GW signal size: $\Delta \theta \sim 10^{-17} \times \left( \mu \text{Hz} / f_{\gw} \right)$ in the frequency band of interest.
In this section, we evaluate these backgrounds and suggest mitigation strategies. 

There are two dominant sources of astrophysical backgrounds that can cause angular position fluctuations. 
The first arises from the presence of starspots or brightness variations on the surface of the star. 
The existence of such a spot does not change the physical location of the center of mass (c.m.) of the star. 
However, it shifts the photometric center and thus the astrometrically inferred position of the star, relative to the projected c.m.~position (see, e.g., \citeR{Makarov:2009nbo}).
As we will see in \sectref{sect:starspots}, the need to mitigate this background strongly motivates the use of non-magnetic white dwarfs as targets for astrometry.
The second background arises from the possible presence of planets and other minor bodies around the star. 
Their motion will cause the c.m.~of the star to wobble due to stellar reflex motion, mimicking a GW signal (indeed, this is an exoplanet search technique).
We estimate this effect in \sectref{sect:planets}.
We also discuss mitigation strategies and cross-checks that could isolate noise sources in \sectref{sect:mitigation}.

\subsection{Starspots}
\label{sect:starspots}
Starspot activity (e.g., appearance, disappearance, and motion) in the 10\,nHz--$\mu$Hz frequency band is a relevant noise for the proposed measurement.
A starspot of radius $r$ with an intensity (i.e., radiated power per stellar surface area) change of $\Delta I$ relative to the nominal intensity of the star will shift the photometric center of the star by, parametrically, $\Delta x \sim (\Delta I/I_0) \times (r^2/R_{\WD})$, where $I_0$ is the mean intensity of the star and $R_{\WD}$ its radius.
The angular deflection caused by this shift is $(\Delta \theta)_{\text{spot}} \sim \Delta x / d_{\WD} \sim (\Delta I/I_0) \times [r^2/(R_{\WD} d_{\WD})] $ where $d_{\WD}$ is the distance between the star and the Earth.  
To see the GW signal above noise, we need $(\Delta \theta)_{\text{spot}} \lesssim h_c$; i.e., $\left[ (\Delta I \, r^2) / ( I_0 R_{\WD}^2 ) \right] R_{\WD} \lesssim h_c d_{\WD} $. 
There are two important takeaways from this expression. 
First, the angular deflection caused by a fixed total fractional luminosity change, $\Delta L / L_0  \sim  (\Delta I A_{\text{spot}} ) / ( I_0 A_{\WD} ) \sim (\Delta I \, r^2) / ( I_0 R_{\WD}^2 ) = \text{const.}$, is \emph{smaller} as the radius $R_{\WD}$ of the star gets smaller.
It is thus advantageous to look at smaller stars, since they can tolerate larger fractional luminosity (and hence, apparent brightness) fluctuations. 
Second, the effect of these fluctuations on the inferred angular position of the star decreases with increasing distance $d_{\WD}$ between the Earth and the star. 
However, one cannot make $d_{\WD}$ arbitrarily large without metrological consequences: a distant star is fainter (higher apparent magnitude), increasing the photon shot noise in the measurement of its angular position: $(\Delta \theta)_{\text{shot}} \propto 1/\sqrt{ N_{\gamma}} \propto d_{\WD}$. 
There is thus a distance trade-off to be made between starspot and photon-statistics noise sources.

In this context, we point out that isolated, hot, non-pulsating, non-magnetic WD are natural candidates for such measurements. 

First, hot WD with $T\gtrsim \text{(few)}\times 10^4 $\,K (the exact threshold depends on the class of WD) are expected to have fully radiative atmospheres, and should therefore lack conventional starspots, as the latter are believed to form owing to magnetic fields inhibiting convective activity in the stellar outskirts when this is the dominant energy transfer process (see, e.g., \citeR[s]{Brinkworth_2013,Strassmeier:2009rga,Calamida_2019}).
Nevertheless, unconventional spots could be present, and/or our understanding of WD surfaces incomplete, and it is therefore useful to have direct empirical handles on the existence of any such spots.
WD time-series photometric data supply one such handle to confirm the absence of starspots.

Since white dwarfs typically have rotational periods of $T_{\WD}\lesssim$ few days [$f_{\WD}\gtrsim 3\, \mu$Hz]~\cite{koester1998search,Hermes:2017nqe} (and some as short as a few hours or less: e.g., \citeR[s]{Kilic_2015,10.1111/j.1365-2966.2004.07538.x}), low-frequency starspot activity in the 10 nHz--$\mu$Hz band is likely to have an observable component that appears at the WD rotational frequency, and that appears as a brightness fluctuation in high-cadence measurements that span a number of rotational periods.
That is, if a starspot that is variable on longer timescales is present on a star that rotates faster than the starspot lifetime and the stellar rotational axis is oriented with respect to the line of sight from Earth in such a way that the location of the spot appears and disappears from view on the rotational period, then the overall brightness of the star will be observed to fluctuate at the rotational period of the star (see, e.g., \citeR{Kilic_2015} for a starspot discovery on a massive WD owing to such a periodic photometric brightness fluctuation).
The absence of such brightness fluctuations thus strongly suggests the absence of lower-frequency starspot activity (i.e., excludes the existence of starspots that survive longer than the rotational period of the WD).%
\footnote{\label{ftnt:diffFreqContent}%
    Note that from the perspective of the low-frequency GW measurement, the higher-frequency rotational-period modulation of the star's photometric position owing to the presence of the spot would be averaged away, leaving only a possible net shift in the inferred photometric position averaged over a number of rotational periods; we care about the existence of spots only to the extent that they \emph{change} on longer timescales, corresponding to the GW periods of interest, as this will cause this average inferred photometric position of the star to drift on relevant timescales.
    } %

A possible exception to the above argument is if the viewing alignment is unlucky given the starspot location on the star, such that the spot does not (fully) disappear from view under WD rotation.
With the tuned exception of a star viewed almost directly down its rotational pole, some reduced brightness variation will however still be present, owing to geometrical effects: i.e., the spot orientation oscillates between being more or less face-on as it rotates on the visible face of the star.
The reduced amplitude of the brightness fluctuation would allow the existence of a larger spot for a given measured upper bound on the brightness variation, but unless tuning of the alignment is assumed, this is usually by a factor of a few.

To gain intuition on the magnitude of these effects consider a spot with $r=R_{\WD}/10$ that is offset by a polar angle of $30^\circ$ from the rotational axis of a star.
In an example case where that rotational axis is tilted toward the observed by $60^{\circ}$, the spot is always visible to the observer.
However, in this case, the light curve of the star still undergoes a fractional brightness variation, as measured over a full rotational period, of approximately the same amplitude as that which would occur if the rotational axis were instead oriented perpendicular to the line of sight, so that the spot becomes fully eclipsed by the star under rotation.
If the rotational axis was however more highly tilted toward the observer, by $80^{\circ}$ instead of $60^\circ$, the fractional brightness variation is reduced by a factor of $\sim 3$ as compared to the original ($30^\circ$/$60^\circ$) orientation.%
\footnote{\label{ftnt:unfavorableAlignment}%
    Particularly unfavorable alignment occurs when the spot position as viewed by the observer rotates so as to cross the stellar diameter line that is perpendicular to the projection of the rotational axis perpendicular to the line of sight.
    } %
On the other hand, if a similar-sized starspot were to instead be located on the equator of a star and the rotational axis were perpendicular to the line of sight, the fractional brightness fluctuation would be maximized and about a factor of $\sim 2$ larger than the original ($30^\circ$/$60^\circ$) orientation.
This demonstrates that an unlikely, tuned, alignment is required to suppress a starspot-induced rotational-periodic brightness variation significantly; such an alignment is unlikely to occur for any one star, and is even more unlikely to occur simultaneously for the multiple stars required to be monitored in the context of the GW measurement.
(See also the mitigation procedures discussed in \sectref{sect:mitigation} that could still be used to ultimately reject any spurious GW signal that might arise from any one star with a missed spot.)

We note that many white dwarfs are non-magnetic and photometrically stable, allowing them to serve as anchors in the Landolt~\cite{1992AJ....104..340L,Landolt_2009}, CALSPEC~\cite{Bohlin_2014,2020AJ....160...21B}, or other~\cite{1995AJ....110.1316B,1996AJ....111.1743B,2019ApJS..241...20N} photometric standards. 
Moreover, there are around 250 known white dwarfs with magnitude less than 19 in the \emph{Kepler/K2} optical bandpass whose (apparent) brightness fluctuations in that band are measured to be smaller than $1\%$ on frequencies larger than $\sim \mu$Hz~\cite{2017MNRAS.468.1946H}; indeed, some of the brighter objects are known to be stable down to the $\sim 10^{-4}$ level in this band~\cite{HermesPrivate}.
Preliminary analysis of ground-based optical photometry of a different sample of hot, isolated, non-magnetic (or, not highly magnetic) white dwarfs~\cite{2019ApJS..241...20N,Calamida_2019} again indicates measured stability at the percent level, although analysis and observation of that sample is ongoing~\cite{Calamida_2019}.
Likewise, short-term (few hour) time-series photometry of various WD has been performed using \emph{Gaia} data~\cite{10.1093/mnras/stw1886}, and upper limits on the variability of many of these WD on these timescales are placed in $\Delta m \sim \text{(few)} \times 10^{-3}$ range, corresponding%
\footnote{\label{ftnt:mag}%
    A magnitude fluctuation of $\Delta m$ corresponds to a fractional brightness fluctuation of $\Delta F/F \approx -(2/5)\ln(10)\Delta m \approx -0.921\Delta m$.
} %
to fractional brightness fluctuations in the 0.1--0.5\,\% range.%
\footnote{\label{ftnt:rotPeriodGaia}%
    A population of variable, pulsating WD is known to rotate with periods ranging from a few to a few tens of hours~\cite{Hermes:2017nqe}, but significantly faster rotation for other WD has also been found for WD with eclipsing spots (see, e.g., \citeR{Kilic_2015} and references therein).
    As the relevant WD here are flux standards, they generally lack time-dependent features that can be easily used to extract a rotational period, and we have been unable to find information on the rotational period of most of these WD.
    We thus caution that these specific \emph{Gaia} data can be used to infer the absence of spot-induced variation only if the WD are rotating sufficiently rapidly to cause a significant change to the spot orientation with respect to the viewer over the duration of the observations.
    } %
Upper bounds on the amplitude of any consistent, short-period ($T \leq 1$\,day) photometric variation for a number of WD observed by TESS~\cite{2022arXiv220103670M} in the optical/near-IR are found to be of order $\sim \text{(few)} \times 10^{-4}$, and any peak-to-peak photometric variation at all on timescales larger than 30\,min to be no larger than one--few percent for it not to have been positively detected.
This sample includes some WD earlier identified as photometric standards~\cite{1995AJ....110.1316B}.
On the other hand, periodic variation at the $\sim 0.1\%$ level or larger in the TESS bandpass has been positively identified in some other WD (e.g., a $\sim 0.1\%$ variation of GD 394~\cite{2020ApJ...897L..31W}, which is a unique system known to exhibit large, 25\% variation in the extreme UV band~\cite{1999AJ....117.2466C,2000ApJ...537..977D}), and such WD would be best avoided as targets for our proposed measurement.

Note also that a number of the WD brightness-fluctuation upper bounds that have been placed are statistically limited by the photon shot noise of the measurements~\cite{2022arXiv220103670M,10.1093/mnras/stw1886}, although systematic effects in the measurement platform can also play a role~\cite{2017MNRAS.468.1946H,2022arXiv220103670M}. 
It is thus possible that with better instrumentation (e.g., larger collection area, greater measurement time, more stable platforms), one may find that the brightness of these stars is even more stable than these bounds, which would suppress the upper bound on this noise floor further. 

Assuming that the brightness is stable to $\sim 10^{-4}$ in line with the best measured stabilities, the angular position of an $R_{\WD} \sim 9000$\,km white dwarf (typical for a carbon-oxygen WD of typical mass $M_{\WD}\sim 0.6 M_{\odot}$) at a distance $d_{\WD} \sim 1$\,kpc would be stable to $\sim 3 \times 10^{-17}$ under the action of a moving starspot of the largest size allowed without violating the brightness fluctuation limit (assuming favorable geometry). 
Indeed, our argument applies equally to the total area of the star covered by spots at any given time, even if there is more than one spot; this is because our argument is based on known observational bounds on the net total brightness fluctuation. 
This would be sufficiently small starspot noise to allow access to a large portion of the GW frequency band of interest.
If additional measurements prove that the brightnesses of these white dwarfs are ultimately even more stable than statistically limited measurements to date have found, it would allow the use of less distant white dwarfs for these measurements (or increased precision).

For the remainder of this paper, we take current limits and adopt a benchmark value of $\sim 10^{-4}$ for brightness stability, and thus consider white dwarfs at a distance $d_{\WD} \sim 1$\,kpc. 
Note that this is conservative in the sense that we are using observational brightness fluctuations to bound the mere presence of starspots; in order for starspot-induced photometric jitter to be a noise source, that jitter would also need to have strong frequency overlap with our band of interest.

\subsection{Planets} 
\label{sect:planets}

A planet orbiting a distant star will cause the c.m.~of the star to wobble periodically around the barycenter of the system.
While this is one of the oldest known (but also one of the more challenging) potential signals for exoplanet detection (see, e.g., \citeR[s]{10.1093/mnras/15.9.228,Benedict:2002tq,Muterspaugh2010,Snellen2018}), it is a background for GW detection.
For this background to not swamp a GW signal, the mass $m_p$ of the planet orbiting a star of mass $M_{\WD}$ at the measurement frequency $\omega_{\gw} = 2 \pi f_{\gw}$ must be small enough to change the position of the star in a direction transverse to the line of sight by less than $\Delta x \sim h_c d_{\WD}$, where $h_c \sim 10^{-17} \times \left( \mu\text{Hz}/ f_{\gw} \right)$ is the target strain sensitivity for this mission in the frequency band $10\,\text{nHz}\lesssim f_{\gw} \lesssim \mu\text{Hz}$.

This requirement yields
\begin{align}
    m_p \lesssim 1.5\times 10^{-8} M_{\odot} \left(\frac{d_{\WD}}{\text{kpc}} \right) \left( \frac{\mu \text{Hz}}{f_{\gw}}\right)^{\frac{1}{3}} \left(\frac{M_{\WD}}{0.6\,M_{\odot}} \right)^{\frac{2}{3}}.
    \label{eq:planetMass}
\end{align}
In \eqref{eq:planetMass}, we assumed that the required strain sensitivity scales as $h_c \sim 10^{-17} \times \left(\mu\text{Hz}/ f_{\gw} \right)$; this leads to a relatively flat ($m_p \propto f_{\gw}^{-1/3}$) dependence on the GW frequency. 
Assuming a mean planetary mass density of $\rho_p \sim 3\,\text{g/cm}^3$, a planet of radius $r_p \gtrsim 1300$\,km would be massive enough to cause a troublesome wobble of the position of the star. 
The size/mass class of problematic objects is thus that of a minor planet or moon. 
Moreover, in order for this motion to be a noise source for this measurement, a planetary body of this class must orbit the star in the GW frequency band of interest, $10\,\text{nHz}\lesssim f_{\gw} \lesssim \mu\text{Hz}$. 
For $M_{\WD}\sim 0.6\,M_{\odot}$, this implies that the planetary body must have an orbital semi-major axis $a$ around the star in the range $0.1\,\text{AU}\lesssim a \lesssim 2\,\text{AU}$, in order for the fundamental orbital period to lie in the measurement band; higher harmonics of the orbital frequency will also enter for eccentric orbits, but are suppressed~(see discussions in, e.g., \citeR[s]{Fedderke:2020yfy,Fedderke:2021kuy}). 

White dwarfs are thus an attractive target from the perspective of this noise source as well. 
The red-giant branch (RGB) and asymptotic-giant branch (AGB) phases of stellar evolution that precede the white dwarf endpoint involve the stellar envelope increasing to radii of up to a few AU (with the achieved size being a function of the mass of the stellar progenitor), as well as significant stellar mass loss in a short time frame ($\sim 10^6$\,yrs)~\cite{Kunitomo_2011,Mustill_2012}. 
$N$-body simulations of the dynamics of multi-planetary systems during these phases indicate that the immediate, initial impact of these events is to clear out planetary bodies from the inner few AU of the system~\cite{Kunitomo_2011,Mustill_2012}.
Closer-in objects are either directly engulfed by the stellar envelope, or undergo orbital decay induced by stellar tidal effects (i.e., the planet raises tides on the expanded stellar envelope, which are damped by the viscosity of the stellar material, leading to orbital energy loss).
Objects that are initially further out avoid this fate, but have their orbital semi-major axes expanded adiabatically by the shallowing of the gravitational potential caused by the mass loss.
However, this does not guarantee that the inner regions of WD systems remain clear of planetary (or smaller) bodies.
Because the AGB mass-loss event `resets' the dynamical age of the planetary system, in systems with multiple planets, subsequent gravitational scattering events and accumulated secular perturbations can later re-populate the inner regions of the WD system with planetary bodies~\cite{10.1093/mnras/stu097,10.1093/mnras/sty446,10.1093/mnrasl/slaa193,10.1111/j.1365-2966.2011.18524.x,2022MNRAS.tmp..472M,Debes_2002}.

Indeed, there is extensive evidence that WD systems are actively polluted by material from asteroidal or planetary bodies that are flung into the inner reaches of the system; e.g., \citeR[s]{Zuckerman:1987qq,1990ApJ...357..216G,2003ApJ...596..477Z,Jura:2003sc,Reach_2005,Becklin_2005,Gansicke:2006kp,Kilic:2007sa,2009ApJ...694..805F,2010ApJ...722..725Z,Debes_2012,2012ApJ...760...26B,Nordhaus2013,Koester2014,Rocchetto2015,Vanderburg:2015xq,Manser2016,Xu__2016,vanSluijs2018,Xu_2018,10.1093/mnras/sty2331,Rebassa-Mansergas2019,Wang_2019,Swan2019,2019Sci...364...66M,10.1093/mnras/staa873,10.1093/mnras/staa3836,10.1093/mnras/staa3940,Dennihy_2020,Melis_2020,Vanderbosch_2020,Xu_2020,10.1093/mnras/stab2949,Guidry_2021,Klein_2021,Lai_2021,Vanderbosch_2021,Farihi2022}.
Metal%
\footnote{\label{ftnt:metal}%
    We utilize the astrophysical definition of metal: any material comprised of elements with an atomic number $Z > 2$. 
    } %
absorption lines are observed in $\mathcal{O}(30\%)$ of WD~\cite{2003ApJ...596..477Z,Becklin_2005,2010ApJ...722..725Z,Debes_2012,Koester2014,Rocchetto2015,Vanderburg:2015xq,Xu__2016,10.1093/mnras/sty446,10.1093/mnras/stab2949,Farihi2022,2022MNRAS.tmp..472M}, and around 3\% of WD are measured to have an excess of infrared (IR) emission for which the consensus model is a radiating accretion dust disk that is heated to IR emission temperatures by the WD radiation field~\cite{1990ApJ...357..216G,Becklin_2005,2009ApJ...694..805F,2012ApJ...760...26B,Debes_2012,Rocchetto2015,Vanderburg:2015xq,Rebassa-Mansergas2019,Farihi2022,2022MNRAS.tmp..472M}.
Additionally, metal emission lines are seen in some fraction of WD, which is evidence for some region of the dust disk having sublimated to gas~\cite{Gansicke:2006kp,Manser2016,2019Sci...364...66M,Dennihy_2020,Melis_2020}, and metal absorption lines are also seen in at least one system that is viewed mostly edge-on~\cite{Xu__2016}.
There are also a number of other systems in which complicated close transits of rocky material have been observed~\cite{Vanderburg:2015xq,Vanderbosch_2020,Guidry_2021,Vanderbosch_2021,Farihi2022}. 
Other systems show an absence of evidence of current active accretion, but exhibit evidence of past accretion events: e.g., Si absorption lines can still be present owing to radiative levitation of Si-bearing material in the WD atmosphere long after the sinking timescale~\cite{Koester2014}.
There is also evidence that some WD systems are host to gas-giant planets~\cite{2019Natur.576...61G,2020Natur.585..363V,Blackman:2021sw}.

The current (simplified) model explaining these observations is that small rocky bodies such as asteroids or minor planets are flung into the interior of the WD system (after the end of the preceding AGB phase) by some perturbing distant mass (perhaps a more massive planet orbiting further out).
These objects approach the WD closer than their Roche limit (typically, $\sim R_{\odot}$), leading to their breakup and eventual grinding and dispersion into a dust disk via collisions or other mechanisms~\cite{1990ApJ...357..216G,Debes_2002,Jura:2003sc,10.1111/j.1365-2966.2011.18524.x,Debes_2012,10.1093/mnras/stu097,10.1093/mnras/sty446,10.1093/mnrasl/slaa193,10.1093/mnras/stab2949,2022MNRAS.tmp..472M}.
As the dust orbits decay, they approach the WD more closely, which can lead to their heating above the gas sublimation temperature, followed by rapid orbital decay and accretion onto the WD owing to the larger gas viscosity~\cite{Koester2014}.

The sinking timescales for metals in the atmospheres of DA-type WD (i.e., those with a hydrogen-dominated atmosphere), of order hours to days~\cite{1986ApJS...61..197P,Manser2016}, are much shorter than the cooling ages of those WD. 
Accretion in these systems must thus be actively ongoing today for the above observational features to be present. 
The largest estimated (instantaneous) accretion rates for rocky-type planetary material on DA WD are in the ballpark of $\sim 10^{9}$\,g/s, or $\sim 10^{-17}M_{\odot}$/yr~\cite{Reach_2005,Kilic:2007sa,Manser2016,10.1093/mnras/staa3836}.
For DB-type WD (helium-dominated atmosphere), the sinking timescale is longer, of order $10^{5}$--$10^{6}$\,yrs~\cite{1986ApJS...61..197P}, but still shorter than the cooling age, and inferred mass accretion rates are larger, maximally around $\sim 10^{11}$\,g/s, or $\sim 10^{-15}M_{\odot}$/yr~
\cite{Manser2016,Xu__2016}.
These rates should however be understood as averaged rates over accretion events occurring over the sinking timescale~\cite{Koester2014}.

Depending on the timescale over which the accretion is assumed to have proceeded, a variety of different estimates show that the total amount of material accreted by such WD may be in the range from the mass of a rocky asteroid of a few tens of kilometers in diameter, up to a minor-planet sized body (but with the accretion in the latter case having occurred over a sizeable fraction of the WD cooling age)~\cite{Jura:2003sc,Reach_2005,Kilic:2007sa,Xu__2016,Wang_2019,Vanderbosch_2020}, with the mass of metals in the photospheres today needed to explain metal absorption lines being toward the lower end of that range.
If one were to assume a steady-state situation with the WD dust disks being replenished at the same rate they accrete onto the WD, then we would expect no more than $\sim 10^{-14}M_{\odot}$ of mass to be added to the disk over a representative 10 year duration of our GW mission.
Because this is much lower than the mass of the problematic planetary bodies we estimated in \eqref{eq:planetMass}, it seems unlikely bodies of a problematic size are undergoing such close approaches to the WD on relevant timescales.
Of course, indications of dust accretion could however be evidence of these systems are host to other (more) stably orbiting perturbing massive bodies of potentially troublesome mass, and one may thus wish to avoid them.

In summary, it has become increasingly clear in the past few decades of observations that WD systems can still be quite dynamical environments, and that there is no guarantee that planets and asteroids are absent from the inner few AU of such systems, notwithstanding an initial inner-system clearing during the AGB phase of stellar evolution.
Overall, it is estimated that around 50\% of WD systems are either accreting rocky material today, or have done so in the past~\cite{Koester2014}.
In order to select for WD that are likely to be less dynamical, or less likely to be host to planetary bodies of a problematic nature, one could thus design a mission to undertake our proposed measurement using evidence of accretion as a veto criterion for the WD.
Because only half of WD are estimated to to have active or past accretion, this would not be too severe a restriction.

Moreover, even were this not a completely successful veto on systems with problematic planetary bodies, in any given WD system we may expect at most perhaps a few minor bodies of a problematic mass to find themselves in the inner stellar system during the WD phase.
With a mission duration $\sim$ 10 years, such minor bodies will prevent the use of that specific stellar system for gravitational-wave detection at the fundamental frequency of the minor body, and at higher orbital-frequency harmonics, within a bandwidth $\sim 1/(10\,\text{yrs})\sim 3$ nHz around each such frequency (assuming the orbit is not significantly perturbed from a two-body Keplerian orbit).
This would remove a relatively small part of the frequency band of interest in any given WD system. 
However, since there are a large number of white dwarfs within $\sim$ kpc distances%
\footnote{\label{ftnt:numberOfWD}%
    Excluding variable, binary, and magnetic WD, and those with known IR excesses (i.e., dust disks), the Montreal White Dwarf Database (MWDD)~\cite{2017ASPC..509....3D,MWDDwebsite} lists $\sim 1.6\times 10^{3}$ WD with $T_{\text{eff}} > 1.5\times 10^{4}\,$K that lie at distances in the range from 1--2\,kpc from Earth, from a total database of $\sim 6.8\times 10^{4}$ WD. 
    These WD have a mean (and modal) mass of $M_{\WD}\sim 0.5M_{\odot}$.
    Moreover, of these $\sim 1.6\times 10^{3}$ WD, roughly $\sim 4\times 10^{2}$ also have $T_{\text{eff}}<2.5\times 10^{4}\,$K.
    The exclusions noted above are all based on the default thresholds for being variable, binary, magnetic, and having an IR excess that are defined in the MWDD search tool, and may not reach the tolerances required for this measurement; we provide these population numbers merely to argue that a large sample of WD in the broadly appropriate class exist, from which appropriate candidates could be selected.
    } %
and many of these white dwarfs are expected to be non-magnetic and photometrically stable, a mission will have a large number of potential WD targets to observe.
It would be rather unlikely for multiple WD systems to exhibit sufficiently large orbiting planetary bodies at the same frequency ($N$-body simulations make clear that the endpoint of the system evolution is sensitive to initial conditions). 
Thus, even if a particular WD system is unusable for a small chunk of the frequency band owing to a planetary disturbance, it should be possible to find other systems where that part of the band should be accessible.

\subsection{Noise Mitigation} 
\label{sect:mitigation}
There are also ways to either mitigate noise or test any putative signal for robustness.
For instance, suppose a single WD is orbited by a planet. 
Because the planetary orbit (and thus the stellar orbit around the barycenter of the system) will in general be elliptical, this will cause a wobble that is not at a single frequency, but rather has higher harmonics that, although suppressed, encode information about the planetary orbit.
It is possible, if the signal is large enough, that this information could be used to fit out the planetary motion to some extent, mitigating the noise.

Moreover, neither star-spot noise nor planetary-wobble noise give rise to a signal exactly degenerate with an astrometric GW signal, in particular because these noises will be specific to the stellar system in question, while the GW signal is common (up to orientation and location effects) to all monitored stellar systems given that we search for the astrometric `Earth term'; see \appref{app:fullSignal}.
In particular, a search could be explicitly designed to look only for a common signal, omitting signals that are dominated by one or two stars.
We have however not quantified here the degree of noise suppression this can achieve.

Nevertheless, it is possible that a sufficiently periodic noise source on any one star could still mislead such an analysis; further robustness checks could however be implemented to mitigate this.
For example, suppose that some common signal is detected when a full collection of $N$ stars is monitored.
To further veto single-system noise perturbations that leak into the common-mode signal, one can form $N$ sub-collections of $(N-1)$ objects each, by sequentially omitting a single star, and repeat the search.
One can then can test for the presence of a putative signal in all $N$ such sub-collections.
Should the signal be absent or reduced (in a statistically significant way) in one particular collection of $(N-1)$ objects, that could be suggestive or diagnostic that the star omitted in forming that particular sub-collection is responsible for the signal in the full collection in a way that is contrary the signal's assumed common nature, perhaps because of some intrinsic noise source. 
Further investigation of that star may then be advised before claiming a positive signal detection.

\section{The Instrument}
\label{sect:instrument}
Having outlined the dominant fundamental (i.e., non-instrumental) noise sources for this measurement, we turn in this section to outlining some of the requirements on the instrument that would be necessary to search for this astrometric deflection signal.

In the frequency band 10\,nHz--$\mu$Hz, the angular fluctuation $\Delta \theta$ of a stellar position induced by a gravitational wave of fixed strain $h_c$ is 
\begin{align}
    \Delta \theta \sim h_c
\end{align}
in a time $T_{\gw} \sim 1/f_{\gw}$, independent of the frequency $f_{\gw}$ of the gravitational wave. 
The expected astrophysical background in gravitational waves yields a characteristic strain that is $h_{c} \sim 10^{-17} \times \left(\mu\text{Hz}/f_{\gw} \right)$, and a number of interesting astrophysical sources with $h_c \sim 10^{-16}$--$10^{-17}$ exist even at $f_{\gw} \sim \mu$Hz~\cite{Sesana:2019vho}. 
We thus consider the parameters necessary to obtain an angular sensitivity in the ballpark of $\Delta \theta \sim h_c \sim (\text{few})\times 10^{-17}$ after integrating for a time $T_{\gw} = 1/f_{\gw} \sim 10^6\,\text{s}$.

The only technology that can plausibly achieve the required extreme angular precision for this mission is space-based stellar interferometry. 
Our interest is aimed at astrometry of photometrically stable, non-magnetic white dwarfs. 
For this purpose, we consider white dwarfs with surface temperatures $T\sim 2\times 10^4$\,K, yielding photons with peak wavelength per Wien's displacement law of $\lambda_{\text{Wien}} \sim 1.4\times10^3\,\text{\AA} \sim 0.14\,\mu\text{m}$, which lies in the far UV.
These are somewhat hotter than typical white dwarfs, but there over a thousand of them within a shell from $\sim 1$--$2$\,kpc from Earth (see footnote \ref{ftnt:numberOfWD}).
The photon number flux density at the Earth from such a source with radius $R_{\WD} \sim 9\times 10^3\,\text{km}$ (typical for a $M_{\WD}\sim 0.6M_{\odot}$ WD) and distance $d_{\WD} \sim \text{kpc}$, peaked at $\lambda_{\text{Wien}} \sim 1.4\times10^3$\,\AA, is roughly%
\footnote{\label{ftnt:fluxEstimate}%
    For the purposes of this estimate, we take the rough approximation that we can treat the incoming flux as all being at the peak (Wien) wavelength of the Planck distribution; that is, we estimate $F_0 \sim (\pi^2/60)T^4 \times (4\pi R_{\WD}^2)/(4\pi d_{\WD}^2) / E_{\gamma}$ and we take $E_{\gamma} \sim 2\pi / \lambda_{\text{Wien}} \sim (2\pi/b_{\text{Wien}}) T \approx 4.97 T$.
    Note that if we instead took $E_\gamma \sim T$, then the flux estimate would increase by a factor of $\approx 4.97$, while the estimates at \eqref[s]{eq:deltaThetaRes0} and (\ref{eq:deltaThetaRes}) would increase by $\sqrt{4.97} \approx 2.2$ if we consistently took $\lambda \sim 2\pi/T$ in \eqref{eq:astrometricError}.
    On the other hand, if we used the wavelength at which the number-flux of photons peaks, that would replace the numerical factor of `4.97' in the above with a numerical factor of `3.92', which is a $\sim$20\% correction.
    The upshot is that the estimates at \eqref[s]{eq:deltaThetaRes0} and (\ref{eq:deltaThetaRes}) should be understood to be uncertain by an \order{3} factor arising from these considerations.
    Alternatively, we could argue that we should take only a slice of the spectrum around the Wien peak.
    Were we to restrict to a bandwidth around the Wien peak of $\pm 10\%$ of the Wien peak wavelength, we would keep a fraction $\sim 0.13$ of the total photon flux, but we would keep using $\lambda \sim \lambda_{\text{Wien}}$ in \eqref{eq:astrometricError}.
    This would again only degrade the estimates at \eqref[s]{eq:deltaThetaRes0} and (\ref{eq:deltaThetaRes}) by a factor of $1/\sqrt{0.13}\sim\mathcal{O}(3)$.
} %
$F_0 \sim 5.6\times 10^2\,\text{m}^{-2}\text{s}^{-1}$.

The angular sensitivity of an interferometer with baseline $B$, collecting area $A$, and interrogation time $\tau$ to such a photon source is
\begin{align}
    \Delta \theta &\sim \frac{\lambda}{B} \frac{1}{\sqrt{F_0 A \tau}} \label{eq:astrometricError}\\
    &\sim 4\times 10^{-15} \times\sqrt{\frac{2 \, \text{m}^2}{A}} \times \left(\frac{\text{1 km}}{B} \right) \times \sqrt{\frac{f_{\gw}}{\mu\text{Hz}}} \nl \qquad\times \left( \frac{\lambda}{0.14 \, \mu\text{m}}\right)  \times \sqrt{\frac{5.6\times 10^2\,  \text{m}^{-2}\text{s}^{-1}}{F_0} } ,\label{eq:deltaThetaRes0}\\
    &\sim 3\times 10^{-17} \times \sqrt{\frac{A_{\text{Hubble}}}{A}}  \times \left(\frac{\text{90 km}}{B} \right) \times \sqrt{\frac{f_{\gw}}{\mu\text{Hz}}}\nl \qquad\times \left( \frac{\lambda}{0.14 \, \mu\text{m}}\right) \times \sqrt{\frac{5.6\times 10^2 \,  \text{m}^{-2}\text{s}^{-1}}{F_0} },
    \label{eq:deltaThetaRes}
\end{align}
where we took $\tau \sim T_{\gw} = 1/f_{\gw}$ in order to compare this to \emph{characteristic} strain (i.e., the strain amplitude detectable given one GW period of observation time).

The $A \sim 2\,\text{m}^2$ collecting area and $B \sim 1\,$km baseline we used at \eqref{eq:deltaThetaRes0} are broadly in line with the parameters of the proposed Stellar Imager mission (see \tabref{tab:spaceInterferometers}).
While SI was never launched, its basic parameters, while ambitious, are believed to be technologically possible.
These parameters however do not quite suffice to access the interesting levels of strain in the band of interest. 

We have therefore also provided a more aggressive fiducial estimate at \eqref{eq:deltaThetaRes} assuming two improvements: 
(a) We have increased the collecting area to match that of the Hubble main mirror $A_{\text{Hubble}} \sim \pi (2.4\,\text{m})^2/4\sim 4.5\,\text{m}^2$~\cite{HubbleWebsite}, which allows a $\sim 50$\% reduction in baseline, all else being held equal.
This is however a relatively modest assumption, and less ambitious than the collection area parameters assumed in a proposal such as SPECS~\cite{SPECS_Spirit,Mather:1998zp} (see \tabref{tab:spaceInterferometers}).
We have also assumed (b)~a~baseline closer to 100\,km, which is around two orders of magnitude larger than that considered for either SI or SPECS (see again \tabref{tab:spaceInterferometers}).
We note however that our astrometric GW measurement needs only a single baseline per star (with two collectors per baseline), and not the full imaging capabilities of SI. 
This puts it more in line with the other proposals listed in \tabref{tab:spaceInterferometers} from the viewpoint of the number of independent spacecraft required to be formation flown.

It is also of importance to note that the interferometric fringe contrast in a stellar interferometer is typically degraded, although not completely lost, for baselines $B$ sufficiently long that the source would be resolvable (in principle) by a (hypothetical) single collector with an aperture equal to the baseline~\cite{BERGER2007576}. 
That is, the interferometric `visibility function' falls off rapidly once imaging resolution of the interferometer, $\delta\Theta \sim \lambda /B$ becomes comparable to the apparent angular size of the disk-like stellar source, $\Theta_{\WD}\sim R_{\WD}/d_{\WD}$.
While this is not a strict limit, in order to avoid this issue entirely, we can impose roughly that $\delta\Theta \gtrsim \Theta_{\WD}$. 
This would restrict $B\lesssim \lambda d_{\WD} /R_{\WD} \sim 5\times 10^2$\,km, given the parameters we have assumed here.
The $B \sim 90\,$km baseline indicated at \eqref{eq:deltaThetaRes} is well below this limit, enabling the interferometer to operate with effectively unsuppressed interferometric fringe contrast.  

Note that for the fiducial parameters we assume at \eqref{eq:deltaThetaRes}, we would achieve our rough target strain sensitivity for $f_{\gw} \lesssim 0.5\,\mu$Hz, covering most of our band of interest.
We stress again however that the target strain of $h_c \sim 10^{-17} \times ( \mu\text{Hz} / f_{\gw})$ is fairly rough order-of-magnitude approximation, and louder sources are expected to exist (see, e.g., Fig.~1 of \citeR{Sesana:2019vho}), which may allow a relaxation of parameters.
There is also clearly mission design trade space here to optimize for collecting area and baseline length, since $h_c \propto ( B d_{\text{collector}} )^{-1}$ where $d_{\text{collector}}$ is the light-collector diameter.
It is likely to be easier and less costly to trade off  collection area for longer baselines, where technologically possible.

We note also for the avoidance of doubt that the mission we have in mind here would be similar in conception to the setup of Stellar Imager with regard to achieving the baselines required.
It would \emph{not} consist of a single satellite with such a large baseline; 
rather, independent satellites would be flown in formation with active metrology of the baseline being conducted in real time and along the same optical paths used to perform the stellar interferometry.

In order for spacecraft-to-spacecraft communications to not result in the loss of any light, the communications optics involved in sending the starlight from each collector either to each other or a common interferometric combiner `hub' would need to possess a Rayleigh range somewhat longer than the baseline. 
The requirements on this communications system would be relatively modest compared to the collector optic: a $d_{\text{comm.}} \sim 15\,$cm diameter optic used to direct light from one spacecraft to the other would have a Rayleigh range $z_{\text{Rayleigh}} = \pi d_{\text{comm.}}^2 / ( 4 \lambda ) \sim 125\,\text{km}$ for the starlight at $\lambda \sim 0.14\,\mu\text{m}$.
The same optic would have a Rayleigh range of only $\sim 17$\,km for a metrology laser operating at $1064\,$nm, but the available metrology laser power of $P_{\text{laser}}\sim 1\,$W (typical for an on-orbit laser system) yields a flux vastly larger than the stellar one: the emitted metrology photon number-flux density is $F_{\text{laser}}\sim 5\times 10^{18}\,\text{s}^{-1}A^{-1}_{\text{comm.}}$ where $A_{\text{comm.}} =\pi d_{\text{comm.}}^2/4$.
Therefore, even over the $\sim 90\,$km baseline discussed above at \eqref{eq:deltaThetaRes}, the loss of some metrology laser power would not present a problem.
To see this, consider that the interferometric accuracy with which the baseline could be measured in this setup over one GW period would be $\Delta B \sim \lambda_{\text{laser}} \lb( F_{\text{laser}} A_{\text{comm.}} T_{\gw} (17\,\text{km}/B)^2 \rb)^{-1/2} \sim 2\times 10^{-12} \lambda_{\text{laser}}$ for $B\sim 90\,\text{km}$ and $f_{\gw} = 1/T_{\gw} = 10^{-6}\,$Hz.
But a fractional error on the measurement of the baseline distance $\Delta B$ generically leads to an angular error of order%
\footnote{\label{ftnt:angleError}%
    More precisely, if $\theta$ is the angle of offset of a target from perpendicular to the baseline, the first-order angular error induced by a baseline measurement error is $|\Delta \theta| \sim |(\Delta B/B) \tan\theta|$.
    We have assumed that $\tan \theta \sim \mathcal{O}(1)$ in the text.
} %
$\Delta \theta \sim \Delta B / B$, so that $\Delta \theta \sim 2\times 10^{-12} \lambda_{\text{laser}}/ B \sim 3\times10^{-23}$, which is well below the intrinsic astrometric error estimate at \eqref{eq:deltaThetaRes}.

One additional challenge of our proposed measurement as compared to some other proposed space-based missions in this class is the requirement to simultaneously measure the astrometric positions of pairs of widely separated WD to be able to construct their relative angular fluctuations to the extreme angular precision required. 
This would of course require one interferometric baseline per WD (i.e., four light collectors for a pair of WD), as well as high-precision, active local metrology of the relative orientations and configurations of the multiple baselines to a level at or exceeding the individual stellar interferometric accuracy.
However, for all the same reasons as we discussed above when we noted that local metrology of the baseline distance itself will easily be vastly in excess of requirements to not limit the intrinsic astrometric measurement precision, we would expect that local metrology of the baseline orientations should not limit the measurement.
It may however require some additional optical elements to establish real-time monitoring of all the light collectors' relative distances and orientations.
An evaluation of the engineering required to achieve this is beyond the scope of this paper.

The ballpark parameters of the instrument we have discussed here are set by requiring the white dwarf to be at $d_{\WD} \sim$ kpc distances. 
This distance requirement arose from ensuring that the photometric jitter from any possible starspots was not too large, which was in turn informed by assuming that the brightness fluctuations of the WD were at the current upper limit of $\sim 10^{-4}$ set by existing measurements.
If improved measurements show that the fractional brightness fluctuations are smaller, we would be able to further limit the size of any possible photometric jitter, and it would be possible to use WD that are closer to the Earth. 
The increased flux from a closer star could either be used to decrease the baseline requirement [$h_c \sim \Delta \theta \propto d_{\WD}/B$] on the interferometer, or make it possible to achieve better angular sensitivity for the same fixed $B$. 
The latter option could enable the instrument to probe gravitational waves in this $\mu$Hz frequency band all the way to the astrophysical noise floor. 

We also note that our proposed measurement has a further advantage in our band with respect to pulsar timing arrays.
PTA sensitivity is limited by the radio dish collecting area (really, the signal-to-noise ratio, or signal power above thermal noise for the collectors) and the number of pulsars.
Our sensitivity is also limited by optical collecting area (really, photon shot noise) and number of WD, but we have an additional handle: we can increase the baseline length to increase sensitivity, keeping photon shot noise and number of WD fixed; see \eqref{eq:astrometricError}.
The main trade off would be on the (modest) communications optics between the satellites (in order to avoid loss of light), and possibly station-keeping challenges.

\section{Other Targets}
\label{sect:otherTargets}
Although we have explored non-magnetic, photometrically stable WD as targets in this work, we also mention that there are other stars that could be plausible targets on the basis of their photometric stability.

For instance, the multi-wavelength photometric noise mitigation strategy proposed in \citeR{2021arXiv211206383K} would be able to bring the absolute in-band photometric jitter in the centroids of stars with Sun-like starspot activity down to the level%
\footnote{\label{ftnt:conversions}%
    A note on this estimate is in order: the quantity called `amplitude', hereinafter $\mathcal{A}_k$, in Figs.~5 and 7 of \citeR{2021arXiv211206383K} is defined~\cite{AKLPrivateCommunication} as $\mathcal{A}_k \equiv |A_k|/N$ for frequency bin $k$, where $A_k$ is the discrete Fourier transform of the windowed time-series data as defined in Eq.~(5) of \citeR{2021arXiv211206383K}.
    We account for (1) the window function amplitude suppression $\xi \sim 0.5$, and (2) the partial common-mode signal cancellation by a factor of $\epsilon \sim 1-0.71 \sim 0.29$ inherent in the noise reduction technique described in \citeR{2021arXiv211206383K} and presented in their Fig.~7.
    Then, roughly estimating $\mathcal{A}_k \sim 10^{-7} R_{\text{star}}$ in our band of interest from the upper right-hand panel of Fig.~7 of \citeR{2021arXiv211206383K}, we estimate that the in-band jitter amplitude around $f\sim \mu$Hz is $\Delta x \sim \sqrt{2} \sqrt{ f T } \mathcal{A}_k \times ( \epsilon \xi )^{-1} \sim 2\times 10^{-5}R_{\text{star}}$.
    } %
of $\Delta x \lesssim 2 \times 10^{-5}R_{\text{star}}$ for frequencies $f\sim \mu$Hz.
To achieve an angular fluctuation $\Delta \theta \sim \Delta x / d_{\text{star}} \sim 10^{-17}$ would then require the star to be at a distance of $d_{\text{star}} \sim 2\times 10^{12}R_{\text{star}}$.
For a Sun-like radius star, this would give a required distance on the order of $d_{\text{star}}\sim 48$\,kpc.
This is roughly the diameter of the Milky Way (MW), so some observable stars should still be available at that distance.

Note that without the mitigation strategy of \citeR{2021arXiv211206383K}, the required distance would be a factor of $\sim 10$ larger, which might present a problem given the paucity or absence of stars in the MW that are that distant.
For a Sun-like star with $T\sim 6\times 10^3\,$K (corresponding to $\lambda_{\text{Wien}}\sim 0.5\,\mu$m) at $d_{\text{star}} \sim 48$\,kpc, the photon flux with a Hubble-sized collecting area would be such that $\delta \theta \sim 10^{-17} \times ( 3.5\times 10^3\,\text{km}/B) \times ( 4.5\,\text{m}^2/A)^{1/2}$ is achievable for $T_{\gw} \sim 10^{6}\,$s (again, without the noise mitigation of \citeR{2021arXiv211206383K}, the baseline required would be a factor of $\sim 10$ larger for the same accuracy).
It might thus be possible to utilize Sun-like stars as alternative targets, but they need to be at distances of $\mathcal{O}(50\,\text{kpc})$ to sufficiently suppress spot-induced jitter, and interferometric baselines of $\mathcal{O}(3\times 10^3\,\text{km})$ are required to meet the strain-sensitivity goal given reasonable collecting areas.

To again avoid the issue of interferometric fringe contrast degradation, we would impose roughly that $B \lesssim \lambda d_{\text{star}} /R_{\text{star}} \sim  10^3$\,km, for these parameters. 
This is now slightly more marginal: the $\sim 3.5\times 10^3\,$km baseline nominally discussed above would result in some reduction of interferometric fringe contrast, and a degradation in angular sensitivity.
In the alternative, one could imagine taking $B \lesssim 10^3\,$km, giving up some angular sensitivity while maintaining the full fringe contrast.
It is not clear which of these two choices is optimal; possible future work on these types of sources would need to explore this point in more detail. 

We note that if these same multi-wavelength noise reduction techniques can be successfully applied to the target WD we have considered more fully in this work, it is possible that WD closer to the Earth than $\sim$ kpc would be used, which would allow either improvement of the angular sensitivity, or a relaxation of mission design parameters.

Furthermore, we note that even for our WD targets, we have limited ourselves to the consideration of WD with temperatures $T_{\WD} \lesssim 2\times 10^{4}$\,K. 
We have done this to ensure that the spectral peak does not lie too far into the UV.
Extreme UV (EUV) optics are at present a technological frontier, but advances here could potentially allow the measurement to utilize hotter WD, which would increase the photon flux ($F_0 \propto T^4$), and shorten the light wavelength, both of which would aid to decrease the astrometric error estimate given at \eqref{eq:astrometricError}, allowing either a relaxation of other measurement parameters or better precision for the same parameters.
Of course, even if the spectral peak of a hotter WD lies in the EUV and is thus unavailable absent such technological advances, hotter sources have increased flux at all wavelengths as compared to cooler sources, including at wavelengths longer than the EUV.
As such, although we may not be able to use the peak wavelength for a hotter WD in the astrometric error estimate given at \eqref{eq:astrometricError}, we could take advantage of the larger photon flux $F_0$ for hotter WD at the wavelengths longer than the EUV, which would allow us to improve the accuracy of the measurement regardless.
In connection with this discussion, we note that UV extinction in the Milky Way is relatively severe~\cite{Valencic:2004mc}, particularly over distances larger than $\sim$ kpc, but is less severe out of the plane of the disk, although fewer hot WD would be expected to exist out of the plane of the disk.
For the purposes of this measurement, we only require distances of $\lesssim$ kpc, so this may not be an issue. 
There may be some trade space here to optimize between temperature and distance, or simply to use hotter sources and gain flux while still restricting to wavelengths longer than the EUV.

Finally, we note that earlier-type (hotter) main sequence stars are also a possible target. 
They are expected to have lower stellar activity levels than Sun-like stars; although recent Kepler observations \cite{10.1093/mnras/stt322} show evidence of rotationally modulated brightness fluctuations of A-type stars that could be caused by starspots, they would still have lower intrinsic astrometric motion than Sun-like stars. 
Both of these kinds of main-sequence stars could be observed at greater distances than WD for a given telescope collector area, which would help to mitigate exoplanet signals.

We note that X-ray interferometry missions have also been proposed~\cite{Welsh:1992eav,Gendreau:2009svd,Uttley:2019ngm,2020SPIE11444E..7WD}.  
A na\"ive check of the photon shot-noise level of the brightest, localized X-ray sources in the sky, indicates that this may be a promising avenue to pursue for GW detection.  
We defer this to future work.

\section{Conclusions}
\label{sect:conclusions}
The detection of gravitational waves in the frequency band $10\,\text{nHz}\lesssim f_{\gw}\lesssim \mu\text{Hz}$ faces a serious challenge from gravity gradient noise from asteroids if conducted using missions with test masses confined to the (inner) Solar System~\cite{Fedderke:2020yfy}.
This motivates the use of distant astrophysical objects as test masses to search for gravitational waves in this band. 
The inability of pulsar timing arrays to reach the desired sensitivity in the upper end of this band necessitates the need for other test bodies. 
In this paper, we have pointed out that extremely high-precision astrometry of a small number of non-magnetic, hot white dwarfs offers a promising way to probe gravitational waves in this frequency band. 

The measured brightness fluctuations of a number of non-magnetic white dwarfs show that starspot activity in these white dwarfs is small enough to suppress the associated photometric jitter below required noise levels.
Moreover, initial clearing of the inner reaches of WD systems of problematic minor bodies and planets that are close enough to cause the center of mass of the star to jitter in the measurement band is expected during the red and asymptotic giant branch phases of their progenitor evolution; however, there is a large and growing body of evidence that dynamical gravitational processes can serve to subsequently re-populate the inner parts of the system with these bodies.
That said, about half of white dwarfs show no evidence of accretion of rocky material perturbed by this kind of subsequent evolution, and these cleaner systems may thus be the ideal targets for which to aim.
Ultimately, although any one stellar system might host a planet(s) that would render that system unusable in some narrow frequency bands around the planetary orbital frequency (and its harmonics, for elliptical orbits), they will not give rise to the common signal among all WD that is characterstic of the GW perturbation, and there are further mitigation strategies that can be employed to veto any such signal.

With white dwarfs at $\sim\text{kpc}$ distances, the astrometric instrument necessary to detect astrophysically interesting gravitational waves would need to be a space-based stellar interferometer, with instrumental parameters comparable to those of the Stellar Imager mission proposed in the 2000s.
Longer interferometric baselines, on the order of a hundred kilometers, are however required.
However, as we operate astrometrically, we require only a single baseline per WD to perform the GW measurement, not the much larger number of baselines that SI needed to achieve imaging capabilities.
That is, since our mission proposal must be able to simultaneously track the relative angle between at least two widely separated stars, we require at least four light collectors for two baselines, but that is far fewer than the 20--30 envisaged for SI.
We would however require a larger number of local active metrology systems to continually monitor the relative orientation of the instrument.
These parameters could be relaxed if a population of white dwarfs whose brightness is more stable than current upper limits are found, as this would guarantee lower starpsot activity, allowing closer WD to be utilized. 
This highlights the need for additional time-series photometric measurements of such white dwarfs.

The possibility of using an instrument akin to Stellar Imager to detect gravitational waves adds to the robust science case for space-based stellar interferometric and/or imaging missions. 
In addition to the many other astrophysical goals of such missions, extremely high precision astrometry may also enable probes of the physics of dark matter. 
Such probes have been explored using pulsar timing arrays (e.g., \citeR[s]{Siegel:2007fz,Seto:2007kj,Baghram:2011is,2012MNRAS.426.1369K,Clark:2015sha,Schutz:2016khr,Dror:2019twh,Chen:2019xse,Ramani:2020hdo,Phillips:2020xmf,Lee:2020wfn,Lee:2021zqw,Gardner:2021ntg,Bramante:2021dyx,Delos:2021rqs}) and some measurements of dark matter and dark-matter substructure have also been performed using astrometry data from the \emph{Gaia} satellite (e.g., \citeR[s]{VanTilburg:2018ykj,Mondino:2020rkn,Mishra-Sharma:2020ynk,Buschmann:2021izy,Gardner:2021ntg}). 
It would be interesting to see what can be learned from a dedicated instrument that focuses on a smaller set of stars but with far greater angular accuracy.

\acknowledgments
We thank Avi Kaplan-Lipkin, Alexander Madurowicz, and Nadia Zakamsa for discussions, and J.~J.~Hermes for useful correspondence. 

S.R.~is supported in part by the U.S.~National Science Foundation (NSF) under Grant No.~PHY-1818899.   
This work was supported by the U.S.~Department of Energy (DOE), Office of Science, National Quantum Information Science Research Centers, Superconducting Quantum Materials and Systems Center (SQMS) under contract No.~DE-AC02-07CH11359. 
S.R.~is also supported by the DOE under a QuantISED grant for MAGIS, and the Simons Investigator Grant No.~827042.
This work was also supported by the Simons Investigator Grant No.~824870, NSF Grant No.~PHY-2014215, DOE HEP QuantISED Award No.~100495, and the Gordon and Betty Moore Foundation Grant No.~GBMF7946.

The work of M.A.F.~was performed in part at the Aspen Center for Physics, which is supported by NSF Grant No.~PHY-1607611.\\

\appendix

\section{Astrometric Deflection of Stellar Position by a Gravitational Wave}
\hypertarget{app:fullSignal}{}
\label{app:fullSignal}
A careful derivation of the astrometric deflection of the inferred position of star owing to the passage of a gravitational wave is given in \citeR{Book:2010pf}, and the signal has been computed elsewhere before (see, e.g., \citeR[s]{Pyne:1995iy,Book:2010pf} and references therein).
We reproduce the final result of \citeR{Book:2010pf} and some of their important discussion points here, for the convenience of the reader.

Suppose the spatial part of the metric in transverse-traceless gauge is $g_{ij} = \eta_{ij} + h_{ij}$ with $|h|\ll 1$; we take the metric signature for the Minkowski background to be mostly plus, and assume that $h_{ij} = h_{ji}$.
Suppose further that $h_{ij}(t,\bm{x}) = \Re\lb[ \mathcal{H}_{ij} \exp[-i\omega_{\gw}(t-\bm{p}\cdot\bm{x}) \rb]$, where $\bm{p}$ is a unit vector that parametrizes the direction of propagation of the GW, and $\mathcal{H}_{ij}=\mathcal{H}_{ji}$ is a constant tensor, in general complex, that we take to satisfy $p^{i} \mathcal{H}_{ij} = 0$.

Consider an orthonormal frame, with directions indexed by $\hat{i}=1,2,3$ and chosen so as to be aligned with the co-ordinate basis if $h\equiv0$, that is parallel transported by an observer located at the origin of the co-ordinate system, $\bm{x}=0$.
Consider a star that is measured to be located at direction $n^{\hat{i}}$ in that frame.
The GW will cause this position to fluctuate: $n^{\hat{i}}(t) = n_0^{\hat{i}} + \delta n^{\hat{i}}(t)$.
To first order in $|h|$, it can be shown that~\cite{Book:2010pf}
\begin{widetext}
\begin{align}
 \delta n^{\hat{i}}(t) &=
 \Re\lb[ \begin{array}{l}
            \dfrac{ e^{-i\omega_{\gw}t} n_0^j n_0^k \mathcal{H}_{jk} }{ 2(1+\bm{p}\cdot\bm{n}_0) } \lb( \begin{array}{l}
                             n_0^i \lb[ 1 - \dfrac{i(2+\bm{p}\cdot\bm{n}_0)}{\omega_{\gw}d(1+\bm{p}\cdot\bm{n}_0)} \lb( 1 - e^{i\omega_{\gw}d(1+\bm{p}\cdot\bm{n}_0)} \rb) \rb] \\[2ex]
                             +          p^i \lb[ 1 - \dfrac{i}{\omega_{\gw}d(1+\bm{p}\cdot\bm{n}_0)} \lb( 1 - e^{i\omega_{\gw}d(1+\bm{p}\cdot\bm{n}_0)} \rb) \rb]
                             \end{array} \rb) \\[6ex]
            - e^{-i\omega_{\gw}t} n_0^j \mathcal{H}_{ij} \lb[ \dfrac{1}{2} - \dfrac{i}{\omega_{\gw}d(1+\bm{p}\cdot\bm{n}_0)} \lb( 1 - e^{i\omega_{\gw}d(1+\bm{p}\cdot\bm{n}_0)} \rb) \rb]
            \end{array} \rb],
    \label{eq:fullSignal}
\end{align}
\end{widetext}
where $t$ is the time of observation at the origin, $d$ is the co-ordinate distance to the source (at this order, this can be consistently replaced with the proper distance), and $\omega_{\gw}$ is the GW angular frequency.
All dot products are consistently at this order taken assuming a spatial metric equal to the identity (note that although $\bm{p}$ has components in the co-ordinate basis and $\bm{n}$ has components in the orthonormal basis, the difference between the co-ordinate and orthonormal basis vectors is $\mathcal{O}(h)$, so this can be ignored at leading order).
Note also that $\bm{n}_0\cdot \bm{\delta n}=0$, so that both $\bm{n}$ and $\bm{n}_0$ are unit vectors to $\mathcal{O}(h)$~\cite{Book:2010pf}.

If $\bm{n}_0^{\hat{i}} \equiv (\cos\phi\sin\theta,\sin\phi\sin\theta,\cos\theta)^{\hat{i}}$, then the angular deflections in this orthonormal co-ordinate frame are given to leading order in $|\mathcal{H}|\ll 1$ by~\cite{Book:2010pf}
\begin{align}
 \delta \theta &= \delta n^{\hat{i}} \hat{\theta}_0^{\hat{i}}, & \bm{\hat{\theta}}_0^{\hat{i}}&\equiv (\cos\phi\cos\theta,\sin\phi\cos\theta,-\sin\theta)^{\hat{i}}; \\
 \delta \phi &= \frac{ \delta n^{\hat{i}} \hat{\phi}_0^{\hat{i}}}{\sin\theta}, & \bm{\hat{\phi}}_0^{\hat{i}}&\equiv (-\sin\phi,\cos\phi,0)^{\hat{i}}.
\end{align}
Finding the fluctuation in the angular separation $\psi(t) \equiv \psi_0 + \delta \psi(t)$ between any two stars $n=(1),(2)$ is also straightforward: at leading order,
\begin{align}
    \cos\psi(t) &= \bm{n}_{(1)}(t) \cdot \bm{n}_{(2)}(t) \\
    \Rightarrow \delta \psi(t) &= - \frac{ \bm{n}_{0,(1)} \cdot \bm{\delta n}_{(2)}(t) +  \bm{n}_{0,(2)} \cdot \bm{\delta n}_{(1)}(t) }{ \sqrt{ 1 - ( \bm{n}_{0,(1)} \cdot \bm{n}_{0,(2)} )^2}},
\end{align}
where, again, all dot products here are taken assuming a spatial metric equal to the identity as they are evaluated in the parallel-transported orthonormal co-ordinate frame carried by the observer.
Note that the apparent discontinuity at $\bm{n}_{0,(1)} \cdot \bm{n}_{0,(2)}=1$ is removable since $\delta\psi(t) \equiv 0$ when the stars are co-located.

In the distant-source limit, $d\omega_{\gw} \gg 1$, only the local (`Earth term') remains in \eqref{eq:fullSignal}~\cite{Book:2010pf}:
\begin{align}
 \delta n^{\hat{i}} &=
 \Re\lb[ (n_0^i+p^i) \dfrac{ e^{-i\omega_{\gw}t} n_0^j n_0^k \mathcal{H}_{jk} }{ 2(1+\bm{p}\cdot\bm{n}_0) }
            - \frac{1}{2} e^{-i\omega_{\gw}t} n_0^j \mathcal{H}_{ij} \rb]\\
            &= \lb[  \dfrac{ (n_0^i+p^i)n_0^k }{ 2(1+\bm{p}\cdot\bm{n}_0) }
            - \frac{1}{2} \delta^{ik}  \rb]n_0^j \times \Re\lb[ e^{-i\omega_{\gw}t}   \mathcal{H}_{jk} \rb],
    \label{eq:fullSignalFarSource}
\end{align}
where at the last step we assumed without loss of generality that $\bm{p}$ and $\bm{n}_0$ are real 3-vectors and we used the symmetry of $\mathcal{H}$.

If we take $\bm{p} = \bm{\hat{z}}$, then the GW can be parametrized up to an overall phase that is degenerate with a time translation by $\mathcal{H}_{ij} = h_{+}^{(0)} \lb( \delta_{i1}\delta_{j1} - \delta_{i2}\delta_{j2} \rb) + h_{\times}^{(0)} e^{-i\alpha} \lb( \delta_{i1}\delta_{j2} + \delta_{i2}\delta_{j1} \rb)$ where $h_{+,\times}^{(0)}$ are the amplitudes of the $+,\times$ polarizations, respectively, and $\alpha$ is a phase.
Then we have angular deflections with respect to the observer's parallel-transported orthonormal co-ordinate frame of
\begin{align}
 \delta \theta &= -\frac{h_{+}^{(0)}}{2} \sin(\theta)\cos(2\phi) \cos(\omega_{\gw} t) \nl
 \quad -\frac{h_{\times}^{(0)}}{2} \sin(\theta) \sin(2\phi) \cos(\omega_{\gw} t + \alpha); \\[2ex]
 \delta \phi &= \frac{h_{+}^{(0)}}{2}  \sin(2\phi) \cos(\omega_{\gw} t)\nl
                \quad - \frac{h_{\times}^{(0)}}{2} \cos(2\phi) \cos(\omega_{\gw} t + \alpha). 
    \label{eq:angleDeflectionsFarSource}
\end{align}
The quadrupolar nature of this result (i.e., terms $\propto \sin2\phi,\ \cos2\phi$)  is manifest.
Moreover, the deflections are of \order{h^{(0)}_{+,\times}}, and oscillate with frequency $f_{\gw}$.
%
\bibliographystyle{JHEP}
\bibliography{references.bib}
%
\end{document}